# Spin torque due to non-uniform Rashba spin orbit effect

Ji Chen [1,a)], Mansoor Bin Abdul Jalil[1,2)], Seng Ghee Tan [1,3)]

[1] *Computational Nanoelectronics and Nano-device Laboratory, Electrical and Computer Engineering Department, National University of Singapore, 4 Engineering Drive 3, Singapore 117576*

[2] *Information Storage Materials Laboratory, Electrical and Computer Engineering Department, National University of Singapore, 4 Engineering Drive 3, Singapore 117576*

[3] *Data Storage Institute, A*STAR (Agency for Science, Technology and Research), DSI Building, 5 Engineering Drive 1, Singapore 117608*

**Abstracts**

Following the early theoretical descriptions of the spin-orbit-induced spin torque [S.G. Tan *et al*., arXiv:0705.3502 (2007); S.G.Tan *et al*., Ann. Phys.326, 207 (2011)], the first experimental observation of such effect was reported by *L. M. Miron et al., Nature Mater, 9, 230 (2010).* We present in this article three additional spin torque terms that arise from the non-uniformity in magnetization space of the Rashba spin-orbit effect. We propose a simple Rashba gradient device which could potentially lower switching current by $n$ orders of magnitude, where large $n$ measures a small magnetization change.

a) Electronic mail: g0800546@nus.edu.sg

**Introduction**

Manipulating the magnetization of ferromagnetic (FM) materials by means of the spin transfer torque (STT) has been theoretically predicted in 1996 **[1,2]**. This prediction has since triggered a series of theoretical and experimental studies in both fundamental and application fields. In the applied physics community, the STT phenomenon can be observed in common nanoscale devices like the spin valve sensors, the magnetic tunnel junction storage cell, as well as in magnetic domain walls **[3-5]**. In fact, STT is now widely used for a magnetization switching process known as the current-induced magnetization switching (CIMS). The adoption of CIMS in the switching operation of memory cells has enhanced the packing density of cross-bar memory, and has significantly moved forward the commercialization schedule of these memories.

Recently it was found that spin orbit coupling (SOC) effect in FM materials could yet be another source of spin torque. The best known example is the Rashba SOC, originating from the structural inversion asymmetry (SIA), which has been found in semiconductor heterostructure **[6-8]** or at the surface of non-magnetic or magnetic metal multilayers**[9,10]**. Rashba spin torque was theoretically predicted independently by Tan et al. **[11]**, Obata et al. **[12]**, and Manchon et al. **[13]** to exist in ferromagnetic materials with Rashba spin orbit coupling. Experimental demonstration of this SOC spin torque effect on the bistable states of a cobalt layer was recently reported in nanosized devices **[14,15]**. The physics of the SOC spin torque is different from the STT in that spin injection would not be needed in the case of SOC spin torque. It suffices to pass charge current directly into the FM material with SOC, the effect of spin polarization and spin transfer takes place within the single FM layer. By contrast, STT system requires a spin polarizing layer (normally a FM layer with hard magnetization) to spin polarize electrical current, and an efficient transport mechanism to inject spin current into a relatively soft FM layer where spin transfer torque could occur within.

SOC spin torque in a uniform FM system with Rashba effect has been well understood. In the event that Rashba constant varies spatially, a similar distribution of the spin torque would follow straightforwardly. In this paper, we will discuss spin torque effects in a magnetic system with non-uniform spin orbit constant in the space of the magnetization. In this system, Rashba constant is a function of the magnetization i.e. $(M)$, as well as its spatial gradient $(\nabla M)$. We consider a Hamiltonian, where care has been taken to ensure its Hermiticity, of a FM structure with Rashba SOC

$$\mathcal{H} = \frac{1}{2m}(p_x^2 + p_y^2) + \frac{\alpha(r)}{2\hbar}(p_x \sigma_y - p_y \sigma_x) + (p_x \sigma_y - p_y \sigma_x)\frac{\alpha(r)}{2\hbar} + J_{sd}\boldsymbol{\sigma} \cdot \boldsymbol{M}(r)$$

(1)

where $m$ is the effective electron mass, $p_i$ is the conduction band electron momentum, $\alpha(r)$ is a spatially dependent Rashba SOC parameter, $\boldsymbol{\sigma} = (\sigma_x, \sigma_y, \sigma_z)$ are the Pauli spin matrices, $J_{sd}$ is the s-d exchange integral and $\boldsymbol{M}(r)$ is the local magnetization due to the localized d-orbitals in the magnetic materials. The Hamiltonian can be written in the following way

$$\mathcal{H} \approx \frac{1}{2m}(\boldsymbol{p} + e\boldsymbol{A})^2 + J_{sd}\boldsymbol{\sigma} \cdot \boldsymbol{M}(r)$$

(2)

where $\boldsymbol{A} = \frac{\alpha(\boldsymbol{r})m}{e\hbar}(-\sigma_y, \sigma_x, 0)$ is a gauge field due solely to the Rashba SOC, which can also be treated as a spin-dependent magnetic vector potential, and has found applications in spintronics based on non-Abelian gauge theory [16-18]. In the presence of local magnetization, it will be instructive to perform a separate local gauge transformation which essentially paves the way for the description of the adiabatic alignment of electron spin along the local magnetization. This transformation process aligns reference spin quantization axis to the local magnetization, resulting in a Hamiltonian in the rotated frame of

$$\mathcal{H}' = U\mathcal{H}U^\dagger \approx \frac{1}{2m}\left(\boldsymbol{p} + e\left[\alpha(\boldsymbol{r})U\boldsymbol{A}U^\dagger - i\frac{\hbar}{e}U\boldsymbol{\nabla}U^\dagger\right]\right)^2 + J_{sd}\sigma_z|M_z(\boldsymbol{r})|$$

$$\equiv \frac{1}{2m}(\boldsymbol{p} + e\boldsymbol{A}^{RSO} + \boldsymbol{A}^{CH})^2 + J_{sd}\sigma_z|M_z(\boldsymbol{r})|, \qquad (3)$$

where the unitary rotation matrix $U = \begin{pmatrix} \cos\theta/2 & \sin\theta/2\, e^{-i\phi} \\ \sin\theta/2\, e^{i\phi} & -\cos\theta/2 \end{pmatrix}$ has been used, $\theta(\boldsymbol{r})$ and $\phi(\boldsymbol{r})$ are respectively, the polar and azimuth orientations of the local magnetization. Now, $\sigma_z$ is the $z$ Pauli spin matrix of the new reference frame, $|M_z(\boldsymbol{r})| = \sqrt{M_x^2 + M_y^2 + M_z^2}$, $\boldsymbol{A}^{RSO} = \alpha(\boldsymbol{r})U\boldsymbol{A}U^\dagger$ and $\boldsymbol{A}^{CH} = -i\frac{\hbar}{e}U\boldsymbol{\nabla}U^\dagger$ are, respectively, the gauge fields due to the Rashba SOC in the rotated frame, and the spatially varying magnetization (domain wall). For better physical clarity, one can regard the gauge field $\boldsymbol{A}^{RSO}$ (vector potential) as representing an effective spin particle generated by the combined presence of Rashba SOC and local magnetization.

Now consider a current propagating through the FM structure with spin orbit coupling. The interaction between the fermion field (electron) and the gauge field (effective spin particle) is given by

$$E_{int} = \int \psi'^\dagger \boldsymbol{A}^{RSO} \cdot \boldsymbol{\nabla}\psi' - (\boldsymbol{\nabla}\psi'^\dagger) \cdot \boldsymbol{A}^{RSO}\psi'\, d^3x, \qquad (4)$$

where $\psi' = U\psi$. Since in the adiabatic system where $J_{sd} \to \infty$, spin is constantly aligned to the local magnetization, and there is no probability of the spin assuming its other eigenstate, one has $\psi' = \begin{pmatrix} c_\uparrow(r) \\ 0 \end{pmatrix}$, and arrives at

$$E_{int} = \int \boldsymbol{a} \cdot (c_\uparrow^\dagger \boldsymbol{\nabla} c_\uparrow) - (c_\uparrow \boldsymbol{\nabla} c_\uparrow^\dagger) \cdot \boldsymbol{a} \; d^3x = \int \boldsymbol{J}_\uparrow \cdot \boldsymbol{a} \; dV, \tag{5}$$

where $\boldsymbol{a}$ is the positive strength of the diagonal matrix component of $\boldsymbol{A}^{RSO}$, and $\boldsymbol{J}_\uparrow$ is equivalent to the electric current density under the adiabatic approximation, $V$ is the volume of the magnetic material. In the spirit of micromagnetic studies, an effective anisotropy magnetic field due to Rashba SOC and local magnetization can now be obtained by taking the energy gradient with respect to the local magnetization, i.e. $\boldsymbol{H}^{\text{eff}} = -\frac{1}{\mu_0} \frac{\delta E_{int}}{\delta \boldsymbol{M}}$ which yields the Rashba-induced anisotropy field of

$$\boldsymbol{H}^{RSO} = \frac{m\alpha}{Me\hbar\mu_0} \boldsymbol{z} \times \boldsymbol{j} + \frac{m}{e\hbar\mu_0} \boldsymbol{z} \cdot (\boldsymbol{j} \times \boldsymbol{n}) \frac{\partial \alpha}{\partial \boldsymbol{M}} + \frac{m}{e\hbar\mu_0} \left( \boldsymbol{\nabla} \cdot \frac{\partial \alpha}{\partial \boldsymbol{\nabla} \boldsymbol{M}} \right) \boldsymbol{z} \cdot (\boldsymbol{n} \times \boldsymbol{j})$$

$$+ \frac{m}{e\hbar\mu_0} \frac{\partial \alpha}{\partial \boldsymbol{\nabla} \boldsymbol{M}} \cdot \boldsymbol{\nabla}(\boldsymbol{z} \cdot (\boldsymbol{n} \times \boldsymbol{j})), \tag{6}$$

where $\boldsymbol{n} = \boldsymbol{M}/M$, $\boldsymbol{z}$ is a z-direction unit vector and $\mu_0 = 4\pi \times 10^{-7} \; TmA^{-1}$ is the vacuum permeability. Note that, to be precise, $\boldsymbol{H}^{\text{eff}} = \boldsymbol{H}^{RSO} + \boldsymbol{H}^{CH}$, where $\boldsymbol{H}^{CH}$ arises from the domain wall effect which is related to $\boldsymbol{A}^{CH}$. However, in this study, our focus is on $\boldsymbol{H}^{RSO}$ only. The first term is coincident with the results already discussed in the previous studies **[11-13]**, while the second, third, and the fourth terms are obtained because of the magnetization effects on the Rashba constant. The second term exists when Rashba strength is varying with the local magnetization, while the third and fourth terms exist when Rashba strength is varying with the spatial gradient of the local magnetization or for simplicity, domain wall.

The dynamics of $M$ can be described by the Landau-Lifshitz-Gilbert (LLG) equation. The total energy of this magnetic system also includes the exchange energy, magnetostatic energy, anisotropy energy, as well as the electron-spin-particle interaction energy which gives rise to $H^{RSO}$. In the low-damping limit, the local magnetization will precess about a total effective field $H$ which can be obtained from the total energy functional, thus $H = H^{exch} + H^{mag} + H^{ani} + H^{RSO}$. Thus, the effective anisotropy field due to the Rashba SOC can be treated as an externally applied magnetic field. The general equation of motion (EOM) for the local magnetization can thus be written as $\frac{dM}{dt} = -\gamma(M \times H)$, where $\gamma$ is the gyromagnetic ratio with units of $mA^{-1}s^{-1}$. For simplicity, the spin torque due to Rashba effect is written as $T_{RSO} = -\gamma(M \times H^{RSO})$ which also gives a simple heuristic picture of the effects of the anisotropy fields on $M$.

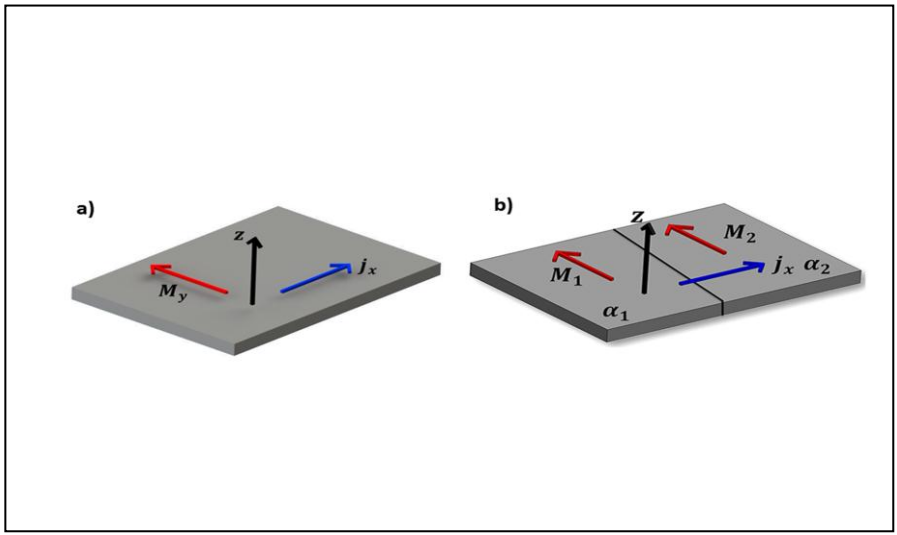

FIG.1. (a) Schematic of a nanoscale device with current fixed along $x$, and the Rashba effective electric field along $z$, perpendicular to the material surface. (b) a Rashba gradient device comprising two devices of type (a). Technology that ensures a maximum $\alpha_1 - \alpha_2$, and a minimum $M_1 - M_2$, would deliver a maximum reduction of switching current density.

In the context of Fig.1(a), the first Rashba anisotropy field term of $\frac{\alpha(r)}{Me\hbar\mu_0}j_x$ can generate motion of $M_z$ or $M_x$ [11-12, 14]; one can verify the above by inspecting $\boldsymbol{T}_{RSO} = -\gamma(\boldsymbol{M} \times \boldsymbol{H}^{RSO})$ with $\boldsymbol{H}^{RSO} = H_y$. The second anisotropy fields of $\frac{m}{e\hbar\mu_0}\boldsymbol{z}\cdot(\boldsymbol{j}\times\boldsymbol{n})\frac{\partial\alpha(r)}{\partial\boldsymbol{M}}$ is related to the $\alpha$ gradient in magnetization space. These $\boldsymbol{H}$ fields can exist in all directions depending on the orientation of $\frac{\partial\alpha}{\partial\boldsymbol{M}}$ (see Table 1). The third term which contains $\left(\boldsymbol{\nabla}\cdot\frac{\partial\alpha}{\partial\boldsymbol{\nabla M}}\right)$ can be described as the flux of the $\alpha$ gradient with domain wall, where $\nabla\boldsymbol{M}$ represents the presence of domain wall. The last term is a specifically domain wall effect of $M_y$. In fact, each component of the effective field depends on $\partial_x M_y$, $\partial_y M_y$, and $\partial_z M_y$, where the first two in-plane terms describe the Neel wall, and the third vertical term describes the Bloch wall. Except for the first anisotropy field which depends on the $\alpha$ constant, all new anisotropy fields depend on the $\alpha$ gradient in different spaces and are independent of local $M$. It is thus reasonable to suggest that these new anisotropy fields could potentially be useful for setting off the dynamics or triggering the switching of local moment with high saturation. We summarize the effects of the Rashba anisotropy fields for device of Fig.1(a). with in-plane $j_x$ as follows:

Table 1. Summary of Rashba anisotropy field under the current density of $j_x$.

| | Rashba anisotropy field expressions ($H$) | Under the effect of current ($j_x$) | $H$ is the physical effect of |
|---|---|---|---|
| 1. | $H = \dfrac{m\alpha}{Me\hbar\mu_0} z \times j$ | $H = (0, \dfrac{m\alpha}{Me\hbar\mu_0} j_x, 0)$ | Rashba constant |
| 2. | $H = \dfrac{m}{e\hbar\mu_0} z \cdot (j \times n) \dfrac{\partial \alpha}{\partial M}$ | $H = \dfrac{mj_x}{e\hbar\mu_0} (\dfrac{\partial \alpha}{\partial M_x}, \dfrac{\partial \alpha}{\partial M_y}, \dfrac{\partial \alpha}{\partial M_z})$ | Rashba gradient with local moment |
| 3. | $H = \dfrac{m}{e\hbar\mu_0} \nabla \cdot \dfrac{\partial \alpha}{\partial \nabla M} z \cdot (n \times j)$ | $H = \dfrac{-mj_x}{e\hbar\mu_0} (\nabla \cdot \dfrac{\partial \alpha}{\partial \nabla M_x}, \nabla \cdot \dfrac{\partial \alpha}{\partial \nabla M_y}, \nabla \cdot \dfrac{\partial \alpha}{\partial \nabla M_z})$ | Divergence of [Rashba gradient with domain wall] |
| 4. | $H = \dfrac{m}{e\hbar\mu_0} \dfrac{\partial \alpha}{\partial \nabla M} \cdot \nabla(z \cdot (n \times j))$ | $H = \dfrac{-mj_x}{M_y e\hbar\mu_0} (\dfrac{\partial \alpha}{\partial \nabla M_x} \cdot \nabla M_y, \dfrac{\partial \alpha}{\partial \nabla M_y} \cdot \nabla M_y, \dfrac{\partial \alpha}{\partial \nabla M_z} \cdot \nabla M_y)$ | [Rashba gradient with domain wall]× [Domain wall] |

One now considers a FM nanoscale structure with an anisotropy field $H^{ani}$ along a particular axis and a current density flowing along the $x$ axis (Fig.1). Critical switching current can be estimated from the simple relation of $H^C = H^{RSO}$, where $H^C$ is the critical switching field. For reference, the switching fields for common ferromagnetic metals like Fe, Ni and Co are, respectively, $H^C_{Fe} = 565\ Oe\ (4.5 \times 10^4 Am^{-1})$, $H^C_{Ni} = 233\ Oe\ (1.85 \times 10^4 Am^{-1})$, and $H^C_{Co} = 7{,}429\ Oe\ (5.91 \times 10^5 Am^{-1})$. Using $\alpha = 10^{-10} eVm$, $M_{Co} = 1.09 \times 10^6\ Am^{-1}$, current density $j_x = 10^7 Acm^{-2}$, one can estimate $H^{RSO} = 6.3 \times 10^4 Am^{-1} \sim 0.079\ T$ for the first Rashba anisotropy field. One can thus estimate that current density $j_x = 10^8 Acm^{-2}$ will be sufficient to trigger moment dynamics.

In the following, we will estimate the required Rashba gradient with local moment ($\dfrac{\partial \alpha}{\partial M_y}$), and the flux of Rashba gradient with domain wall ($\nabla \cdot \dfrac{\partial \alpha}{\partial \nabla M_y}$), to generate a $H^{RSO} = 5.91 \times 10^5 Am^{-1}$ which matches $H^C_{Co}$, for current density of $j_x = 10^7 Acm^{-2}$.

Numerical estimate shows a required value of $\approx 8.6 \times 10^{-16} \frac{eVm}{Am^{-1}}$ for the three effects of $\frac{\partial \alpha}{\partial M_y}$, $\nabla \cdot \frac{\partial \alpha}{\partial \nabla M_y}$, or $\frac{1}{M_y}\frac{\partial \alpha}{\partial \nabla M_y}\cdot \nabla M_y$, all shares the dimension of $\frac{eVm}{Am^{-1}}$. We propose to realize the effect of $\frac{\partial \alpha}{\partial M_y}$ by having two devices of Fig.1(a) bordering each other as shown in Fig.1(b), which we call a Rashba gradient device. Reduction of switching current density can be achieved by maximizing $\Delta \alpha = \alpha_1 - \alpha_2$, and minimizing $\Delta M = M_1 - M_2$. One crude example is to have device 1 of Oxide/Co/Pt and device 2 of Oxide/Co/Cu with no Rashba effect, where $\Delta \alpha = \alpha_1$, and $\Delta M$ can be minimized with precise fabrication. Now by controlling $\Delta M$ between the two devices to $\frac{\Delta M}{M} = 10^{-n}$, switching current of the Rashba gradient device can be reduced by $j_{gc} = j_c \frac{\Delta M}{M} = j_c \times 10^{-n}$. Thus the device can be designed to switch at low current even in material with high $M$.

The physical significance of $\nabla \cdot \frac{\partial \alpha}{\partial \nabla M_y}$ and $\frac{1}{M_y}\frac{\partial \alpha}{\partial \nabla M_y}\cdot \nabla M_y$ are still relatively abstracts and it remains unclear how these domain wall related effects can be measured or detected experimentally. We think it not suitable at this stage to discuss device implementation based on the spin torque of terms 3 and 4 which rely heavily on domain wall configurations. We, however, remark that one can visualize $\nabla \cdot \frac{\partial \alpha}{\partial \nabla M_y}$ better by considering a simple domain wall of $M_y$ varying along distance $x$ only, in which case, the above simply reduces to $\partial_x \frac{\partial \alpha}{\partial \partial_x M_y}$. It is nonetheless clear and useful to note that if domain wall is eliminated from the device, both domain wall related spin torques will not manifest.

In summary, we have proposed in this paper three additional spin orbit gradient spin torques and discussed the implementation of the Rashba gradient spin torque.

Rasbha gradient with local moment effect has the potential to lower switching current significantly with proper implementation. The domain wall effects are less clear in terms of how they can be harnessed, but we believe the dimension $\frac{eVm}{Am^{-1}}$ has important physical significance here, in the same way it has been used to lower the switching current with respect to the Rashba gradient with local moment. In general, we remain optimistic that the challenges related to the domain wall can be overcome in the near future and the estimate we provided above will make instant practical sense.